\documentclass[twocolumn]{aastex62}

\newcommand{\titlerunning}[1]{\shorttitle{#1}}
\newcommand{\authorrunning}[1]{\shortauthors{#1}}

\newcommand*\inst[1]{\unskip\hbox{\@textsuperscript{\normalfont$#1$}}}

\newcommand*\institute[1]{
  \begingroup
    \let\and\relax
    \renewcommand*\inst[1]{}%
    \renewcommand*\thanks[1]{}%
    \renewcommand*\email[1]{}%
  \endgroup
  \newcommand{\institutions}{#1}
}%

\let\oldarcsec\arcsec
\renewcommand\arcsec{\oldarcsec\xspace}%

\usepackage{latexsym}		
\usepackage{graphicx}		
\usepackage{rotating}		
\usepackage{natbib}  
\usepackage{savesym}
\usepackage{amssymb}
\usepackage{amsmath}
\usepackage{morefloats}
\savesymbol{doublespace}
\usepackage{xspace}
\usepackage{color}
\usepackage{mdframed}
\usepackage{url}
\usepackage{subfigure}
\usepackage{grffile}
\usepackage{import}
\usepackage[utf8]{inputenc}
\usepackage{booktabs}
\usepackage{fancyhdr}

\usepackage[hang,flushmargin]{footmisc}
\usepackage{ifpdf}

\newcommand{\msun}{\ensuremath{\mathrm{M}_{\odot}}\xspace}			

\newcommand{\lsun}{\ensuremath{\mathrm{L}_{\odot}}\xspace}			

\newcommand{\methanol}{\ensuremath{\textrm{CH}_3\textrm{OH}}\xspace}

\newcommand{\water}{H$_{2}$O\xspace}		

\newcommand{\kms}{\textrm{km~s}\ensuremath{^{-1}}\xspace}	

\newcommand{\ammonia}{NH\ensuremath{_3}\xspace}

\newcommand{\perbeam}{\ensuremath{\textrm{beam}^{-1}}\xspace}

\def\eqref#1{Equation \ref{#1}}

\newenvironment{rotatepage}
{}{}

\newcommand{\florida}{\affiliation{\it{Department of Astronomy, University of Florida, PO Box 112055, USA}}}
\newcommand{\nraojansky}{\affiliation{\it{Jansky fellow of the National Radio Astronomy Observatory, Socorro, NM 87801 USA }}}
\newcommand{\radboud}{\affiliation{\it{Department of Astrophysics/IMAPP, Radboud University Nijmegen, PO Box 9010, 6500 GL Nijmegen, the Netherlands}}}
\newcommand{\allegro}{\affiliation{\it{ALLEGRO/Leiden Observatory, Leiden University, PO Box 9513, 2300 RA Leiden, the Netherlands}}}

\begin{document}
\title{First detection of CS masers around a high-mass young stellar object, W51 e2e}
\titlerunning{CS masers}
\authorrunning{Ginsburg}

\author[0000-0001-6431-9633]{Adam Ginsburg}
\florida
\nraojansky
\author{Ciriaco Goddi}
\allegro
\radboud

\begin{abstract}
We report the discovery of maser emission in the two lowest rotational
transitions of CS toward the high-mass protostar W51 e2e with ALMA and the JVLA.  
The masers from CS J=1-0 and J=2-1 are neither spatially nor spectrally
coincident (they are separated by $\sim150$ AU and $\sim30$ \kms), but both
appear to come from the base of the blueshifted outflow from this source.
These CS masers join a growing list of rarely-detected maser transitions that
may trace a unique phase in the formation of high-mass protostars.
\end{abstract}

\section{Introduction}

Because of their compactness, high brightness, and ubiquity, masers are
unique diagnostic probes of the early stages of star formation.
Long baseline interferometric studies
\citep[e.g.,][]{Matthews2010a,Goddi2011c,Moscadelli2014a,Moscadelli2016a} 
have
revealed that
maser emission lines can trace accretion structures, shocks in outflows, and
disk winds in the circumstellar environments of protostars.

While some chemical species, specifically \methanol, \water, and OH, are
detected in hundreds of star-forming regions across the Galaxy, others are more
rare, such as SiO (around young stars; SiO masers are common in evolved stars),
H$_2$CO, and NH$_3$ \citep[e.g.,][]{Goddi2009b,Cho2016a,Ginsburg2015a,Goddi2015a,Mills2018b}.  
The
rarity of this latter group of masers is so far unexplained, though since they
have only been detected toward high-mass star-forming regions
\citep{Araya2015a,Wilson1993a,Hofner1994a,Cordiner2016a}, they are likely to
be pumped by the strong radiation only present around high-mass protostars.

The rotational transitions of CS have been theorized to mase in some
environments \citep{Schoenberg1988a}, and \citet{Highberger2000a} reported a
possible maser from the CS v=1 J=3-2 transition in IRC+10216, but no masers
have been observed in the rotational transitions of the vibrational
ground-state of CS.

The W51 A high-mass star-forming region (at a distance of 5.4 kpc -
\citealt{Sato2010a}) is one of the most massive and luminous ($\sim10^7$~\lsun)
in the Galaxy \citep{Ginsburg2017b}, and it is host to three high-mass
protostellar systems with $>100$ \msun hot cores, W51 IRS2/North, W51 e2e, and W51 e8
\citep{Ginsburg2017a}.  Among the known high-mass protostars exhibiting maser
emission, W51 IRS2/North stands out as host to some of the rarest maser transitions.
It powers a large number ($>$ 20) of rare \ammonia maser lines as well as rare
SiO masers \citep{Henkel2013a,Goddi2015a,Hasegawa1986a,Eisner2002a}.  W51 e2e and e8, on
the other hand, are more typical high-mass star-forming region in terms of
their maser emission, exhibiting only \methanol, \water, and OH masers
\citep{Goddi2016a}.

We have observed all three hot cores in the W51 A region in the two lowest
transitions of CS, J=1-0 and J=2-1.  We report here the first detection of
maser emission from CS $v=0$ in both the J=1-0 and J=2-1 transitions in
W51 e2e and nondetections in the other two hot cores.

\section{Observations}
\label{sec:observations}
We report observations from three different observing programs: ALMA
2013.1.01596.S \citep{Goddi2018a} observed band 6 (1 mm, SiO J=5-4),
2017.1.00293.S observed Band 3 (3 mm, CS J=2-1), and the Karl G. Jansky Very
Large Array (VLA) program VLA 16B-202 observed Q-band (7 mm, CS J=1-0).  The
ALMA data were taken in long-baseline configurations, and the VLA data were
taken in the most extended configuration (A).

For the ALMA data, both at 1 mm and 3 mm, we use the pipeline-produced
calibrated data for the emission line maps.  The Band 6 (1 mm) continuum data
were self-calibrated as described in \citet{Goddi2018a}, using  nine iterations
of phase-only self-calibration; the solutions were only used for the continuum
data.  

The VLA data were calibrated with the EVLA
pipeline\footnote{\url{https://science.nrao.edu/facilities/vla/data-processing/pipeline}}
with radio frequency interference (RFI) flagging and Hanning smoothing disabled
to avoid flagging bright lines.

We imaged the CS J=2-1 line at 97.980953 GHz with robust 0.5 weighting with 3 \kms
channels, resulting in a beam $0.067\arcsec\times0.043\arcsec$, PA=$-44.5^\circ$,
and sensitivity $\sigma_{rms} = 0.59$ mJy \perbeam (26 K).  The CS J=1-0
line at 48.990955 GHz was included in a continuum band with 6 \kms resolution, and we
imaged it with robust 0 weighting, resulting in a beam
$0.043\arcsec\times0.037\arcsec$, PA=$-64.8^\circ$ and
sensitivity 1.3 mJy \perbeam (420 K).  We also imaged the CS v=1
J=1-0 line
at the same resolution with a sensitivity of $\sigma_{rms} = 1.1$ mJy \perbeam (350 K),
but did not detect it.
The SiO J=5-4 and J=2-1 lines and band 6 (1 mm) continuum were imaged with
robust 0.5 weighting with beam sizes $0.041\arcsec\times0.035\arcsec$,
PA=-44.8$^\circ$ and $0.079\arcsec\times0.053\arcsec$, PA=-42.8$^\circ$,
respectively.

The VLA and ALMA data used the same phase calibrator, J1922+1530, with
coordinates that differ by 0.003\arcsec mostly in RA.  Both the VLA and ALMA
positions are offset from the SIMBAD position by about 0.0015\arcsec.  The VLA
measurement set data incorrectly report the stored coordinate for J1922+1530 as
being in the FK5 system, while the ALMA measurement sets correctly report the
stored coordinate as being in ICRS.  The difference between the ICRS and FK5
system is about 0.03\arcsec at this position, close to the beam size, and
therefore is highly relevant for these data.  We corrected the images for the
coordinate system offset (0.03\arcsec), but did not correct for the
$\approx0.003\arcsec$ discrepancy in calibrator position, so our systematic
pointing error is approximately 0.003\arcsec (16 AU).

We report a check on the velocity frame and a search for variability in the
Appendices.

\section{Results}
\begin{figure*}[htp]
    \includegraphics[width=\textwidth]{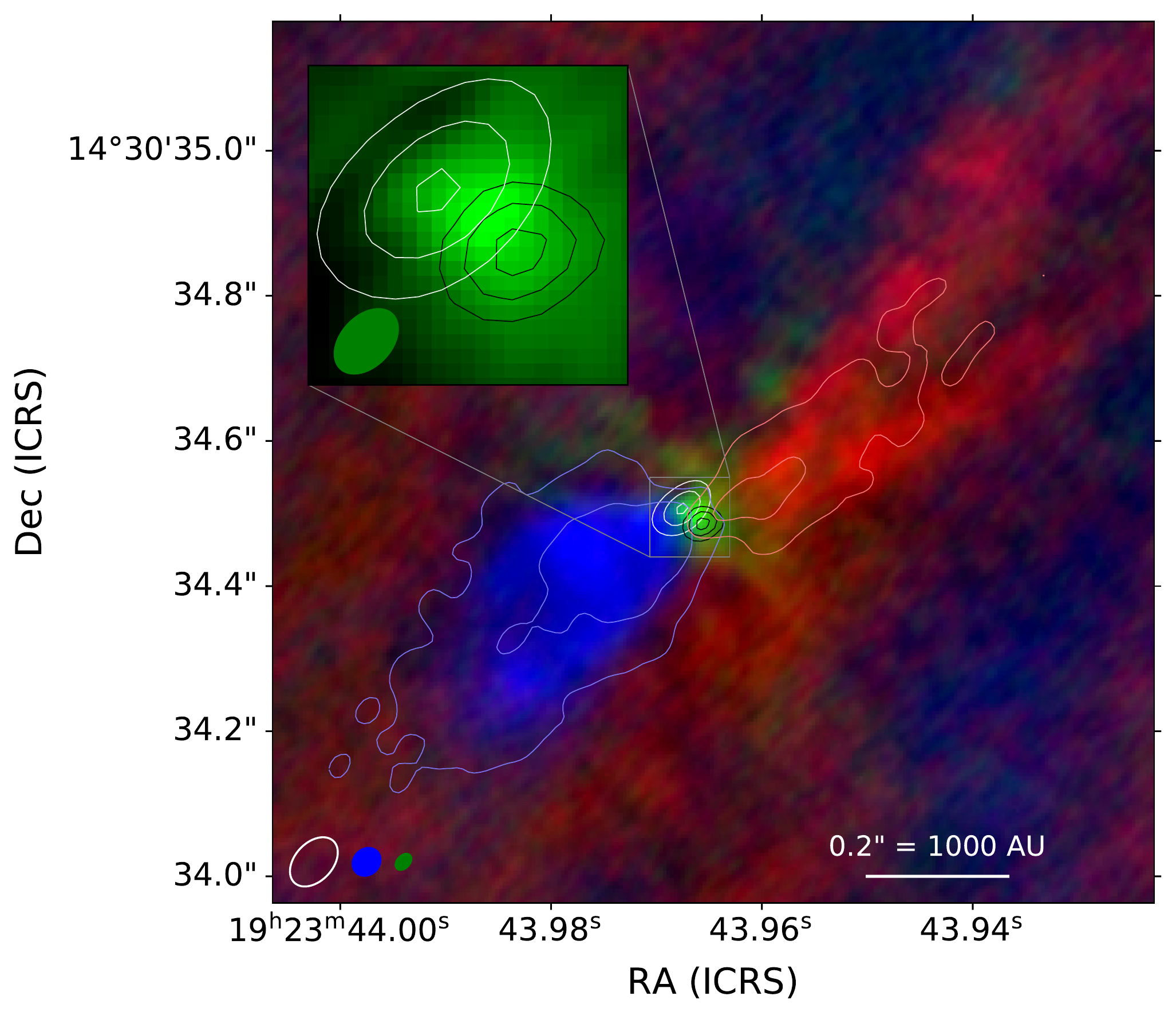}
    \caption{An image of the SiO J=5-4 outflow (red and blue corresponding to
    red and blueshifted emission integrated over 74-118 \kms and -32 to 55
    \kms, respectively) with 1 mm continuum shown in green.  The contours show
    integrated SiO J=2-1 over the same ranges in red (0.05, 0.1 K \kms) and
    blue (0.1, 0.2 K \kms), the CS 2-1
    maser in white (2000, 4000, 6000 K peak intensity at 34.5 \kms),
    and the CS 1-0 maser in black (2000, 4000, 6000 K peak intensity at
    $\sim65.5$ \kms).  For each of the masers, the contours effectively
    indicate the synthesized beam size.  The white, blue, and green
    ellipses in the bottom left show the synthesized beam sizes of the SiO
    J=2-1, SiO J=5-4, and 1 mm continuum images, respectively.
    The inset image shows only the continuum and the maser contours,
    but is otherwise identical.
    }
    \label{fig:overlay}
\end{figure*}

\begin{figure}[htp]
\includegraphics[width=0.45\textwidth]{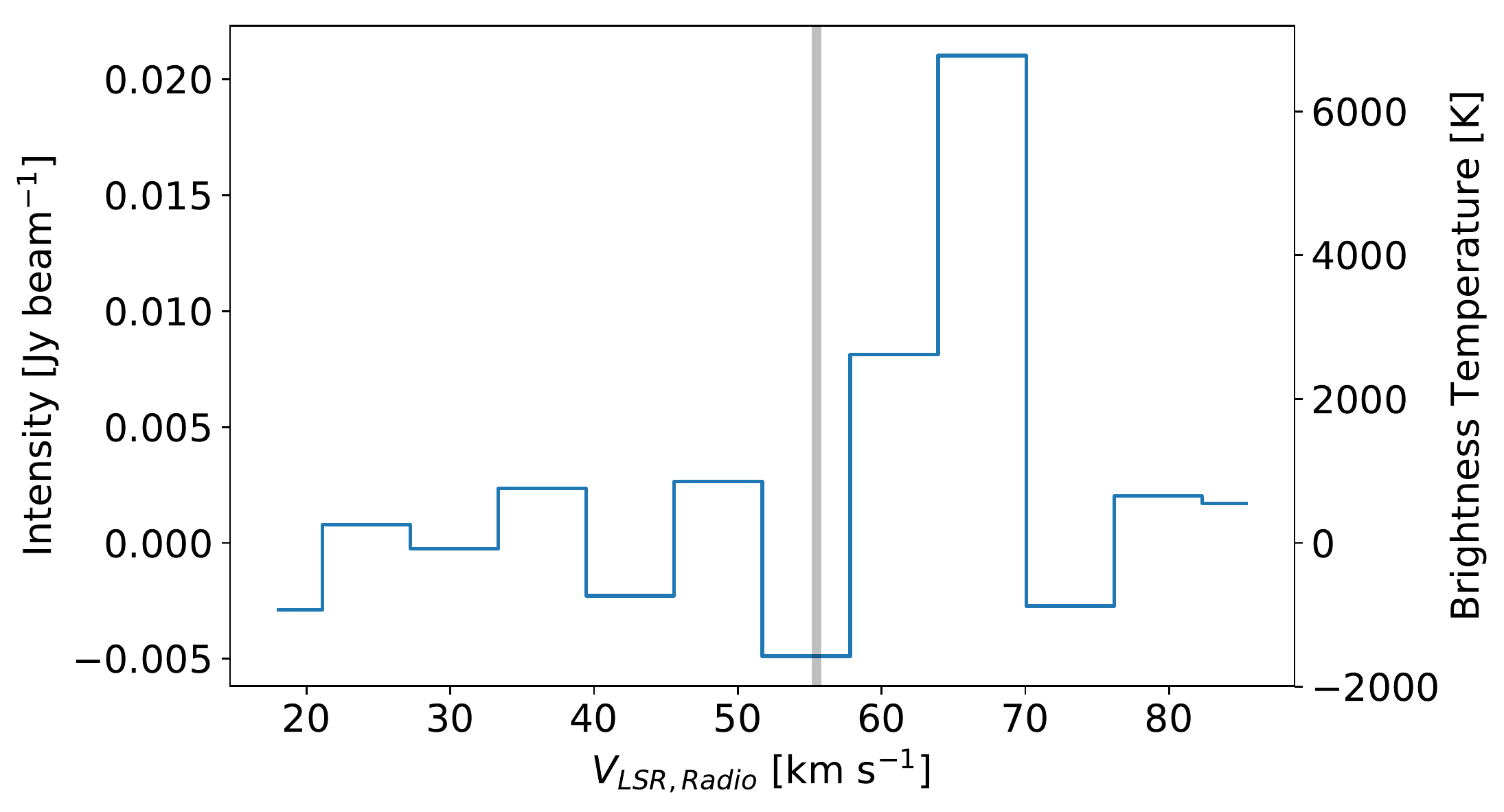}
\includegraphics[width=0.45\textwidth]{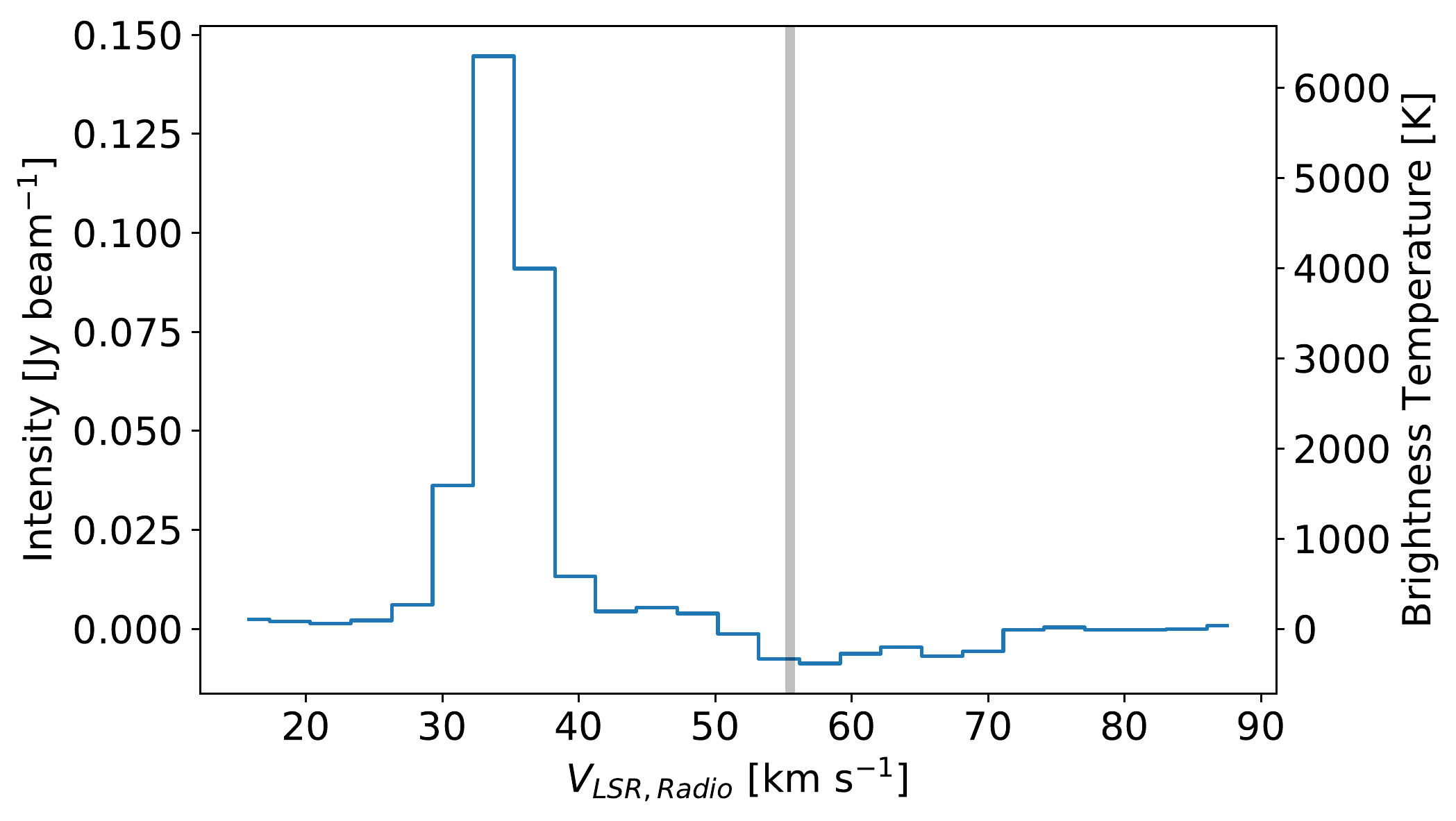}
\caption{Plots of the (a) CS J=1-0 and (b) CS J=2-1 spectra toward their
respective peak locations shown in Figure \ref{fig:overlay}.
The vertical line shows the estimated velocity of W51 e2e from the NH$_3$
inversion lines at 56.4 \kms \citep[][Table 4]{Goddi2016a}.
}
\label{fig:spectra}
\end{figure}

We report the detection of two transitions of CS, J=2-1 and J=1-0, with
peak brightness $T_{B,max}\approx7000$ K.
These lines peak at different locations both spatially (Figure
\ref{fig:overlay}) and spectrally (Figure \ref{fig:spectra}).
We report Gaussian fit parameters to the line profiles and to the peak intensity
images in Table \ref{tab:linepars}; the errors reported in the table are fit
errors assuming no correlation between the parameters, and they do not include
the systematic pointing uncertainty noted in Section \ref{sec:observations}.

\begin{figure*}
    \includegraphics[]{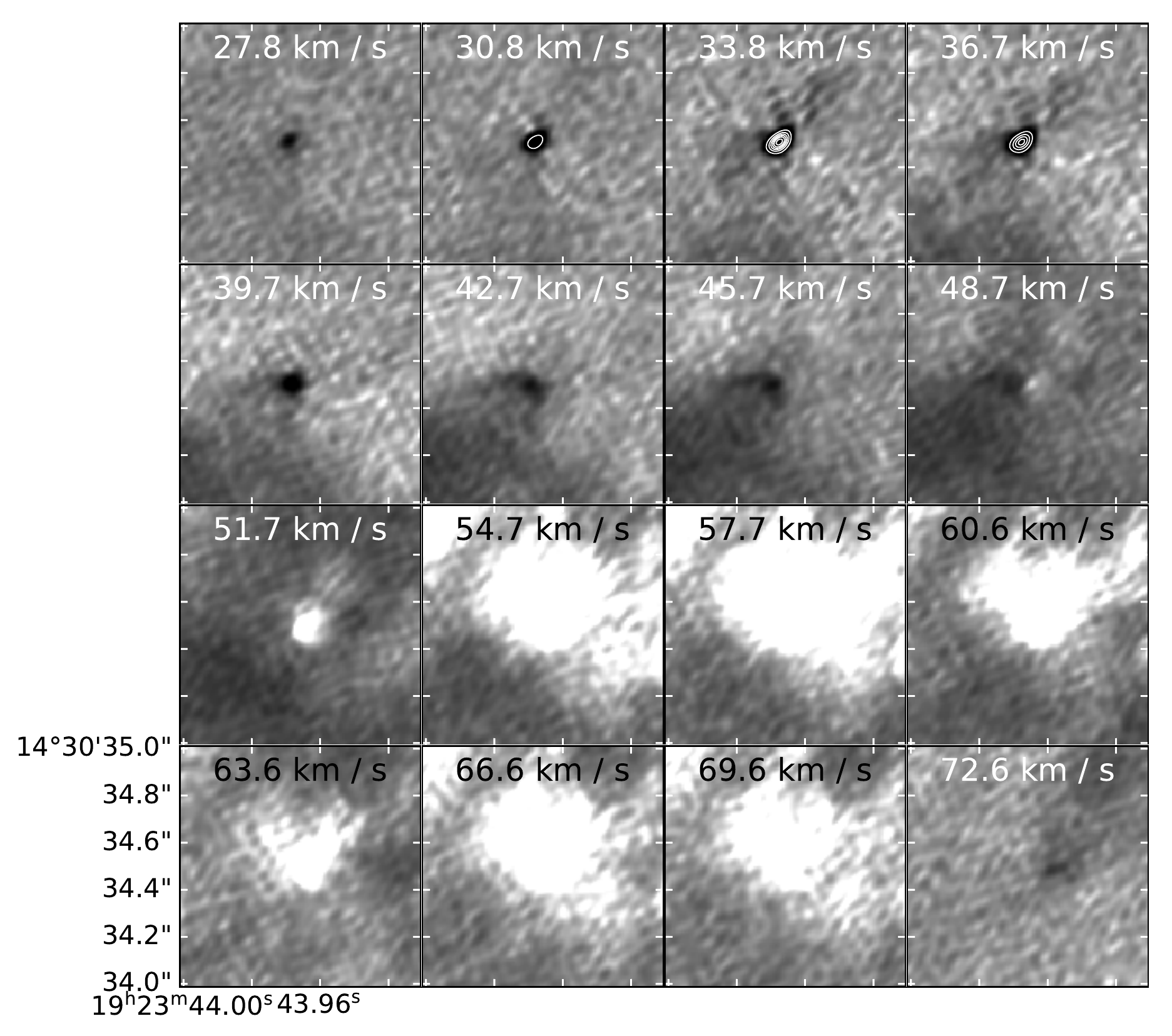}
    \caption{Continuum-subtracted channel maps of the CS J=2-1 transition.
    Channels are 3 \kms wide.  Darker colors indicate higher intensity.  White
    contours are overlaid at [0.02,0.04,0.06,0.08,0.1,0.125] Jy beam$^{-1}$.
    Absorption from the foreground against the continuum is shown as negative
    intensity (white) from 50-70 \kms.  The redshifted outflow is faint but
    still detected at ${\sim72.5~\kms}$.
    }
    \label{fig:channelmaps}
\end{figure*}

The CS J=2-1 line is observed in the blueshifted component of the
previously-detected SiO outflow \citep{Goddi2018a}.
It has both a compact component, which we will refer to as the CS 2-1 maser,
and an extended component that traces the inner envelope of the blueshifted
outflow.  This extended component can be seen more clearly in the 42 \kms
channel of the channel maps (Figure \ref{fig:channelmaps}).  The maser
peak is located at the base of the blueshifted outflow (Figure
\ref{fig:overlay}) at a velocity of $v_{lsr}=34.50\pm0.07$ \kms.

The CS J=1-0 line is only seen as a single spatial component centered
at approximately $v_{lsr}=64 \pm 6$ \kms.  It is centered closer to the
central continuum source, but still slightly toward the blueshifted outflow.
The CS J=1-0 and J=2-1 masers are offset by $0.036\arcsec \pm 0.12$\arcsec
($190\pm60$ AU).

\begin{table*}[htp]
\centering
\caption{Line Fit Parameters}
\begin{tabular}{llllll}
    \label{tab:observations}
Line Name & Amplitude & $v_{LSR}$ & $\sigma_{FWHM}$ & RA (ICRS) & Dec (ICRS) \\
          &         (K) &      (\kms) &            (\kms) & ($\deg$)    & ($\deg$) \\
\hline
CS J=1-0 &      6800                  &$      64 \pm        6$     &$       7.3 \pm        1.2$ &$    290.9331902 \pm       0.0000025$ &$     14.5095795 \pm       0.0000021$ \\
CS J=2-1 &$      6700 \pm        200$ &$     34.50 \pm       0.07$ &$      5.27 \pm       0.17$ &$    290.9331982 \pm       0.0000025$ &$     14.5095852 \pm       0.0000021$ \\
\hline
\end{tabular}
\label{tab:linepars}
\par
\end{table*}

\section{Analysis}
The peak brightness of the observed lines is $\sim6800$ K at $\sim0.07\arcsec$
(350 AU at 3 mm) and $\sim0.04\arcsec$ (200 AU at 7 mm) resolution.  It is
unlikely that the molecules are in thermal equilibrium at $T_K \geq 6000$ K,
since such high
temperatures would more likely result in dissociation of the molecules
\citep[e.g.,][]{Pattillo2018a}.

At the observed spectral resolution of 3 \kms, the CS J=2-1 line is marginally
resolved with ${\mathrm{FWHM} = 5.25~\kms}$ (${\sigma=2.2~\kms}$), and at 6
\kms resolution, the CS J=1-0 line is unresolved.  The thermal line width of CS
at 6000 K is ${\mathrm{FWHM} = 2.7~\kms}$ (${\sigma=1.1~\kms}$),
smaller than the measured width of CS J=2-1 and smaller than the resolution
of the CS J=1-0 data. 
Lines narrower than their thermal width are a clear indication of maser
emission \citep{Elitzur1982a}, but our observations have too coarse resolution
to identify such narrow widths if they are present.

The high observed brightness temperature is partial evidence that these
transitions are masing.  The locations of the emission features provides
additional and definitive evidence that they are masing.  The emission peaks
are at clearly different velocities and they may be from different spatial
locations (separated by about 190 AU).  They are therefore not emitted by the
same material.  This velocity difference rules out a thermal origin for either
transition, since thermal lines should have comparable brightness in both
transitions at similar velocities.

The separation between the two maser spots in position and velocity hints
that the masers could be produced at the extreme ends of a disk orbiting
a central protostar.  
If we assume the masers trace orbiting material, their velocity and spatial
separation can be used to infer the central source mass.  At a separation of 30
\kms and 190 AU, assuming the masers come from opposite ends of a disk such
that $v_{orb}=15\pm1$~\kms and $R_{disk}=95\pm30$~AU, the implied contained mass is
$M=24_{-10}^{+12}$~\msun.  This mass is comparable to that suggested by
\citet{Ginsburg2017a} and \citet{Goddi2018a} based on luminosity estimates.
However, the velocity separation also hints that these masers may be produced
in the outflow rather than in a disk.  In that case, the orbital analysis
is not valid.

Our data also include the W51 IRS2/North and W51 e8 high-mass young stellar
objects (HMYSOs) \citep{Ginsburg2017a}.  No compact and bright CS emission
sources were detected
toward either of these HMYSO candidates, with a peak CS 2-1 flux limit ${S_{98
\mathrm{GHz}} \leq 15 \mathrm{mJy~\perbeam}}$, an order of magnitude fainter
than the peaks seen in W51 e2e.  However, plentiful extended CS 2-1 emission is
seen around each of the HMYSOs.  This difference indicates that there is
something unique about the chemistry, geometry, or excitation in the W51 e2e
that drives these particular maser transitions.

The detection of CS 1-0 emission at $\sim65$ \kms is actually quite surprising,
since W51 e2e is deeply embedded in a high column density molecular medium that
covers this velocity.  
Since the maser is detected, there must either be a hole
in the cloud through which we are seeing the compact emission, or the
foreground cloud's CS
is highly excited and has a minimal ground-state population at 65 \kms. 
The J=2-1 transition at 34.5 \kms is not
subject to this concern, since there is no foreground material at that
velocity, but it is possible that a 2-1 maser exists near 65 \kms and is fully
absorbed by foreground cloud material.

Masers from CS have been predicted in the atmospheres of evolved stars.
\citet{Schoenberg1988a}
used an expanding shell model appropriate to evolved stars in which
temperature, density, and velocity are all decreasing with radius corresponding
to some mass loss rate and terminal wind velocity. They predict masing in the
CS J=1-0 and J=2-1 lines under different conditions, though the masing is
fairly weak (factors of a few).  In these models, the stellar infrared
radiation drives the maser.  The similarity between the atmospheres
of evolved stars and the innermost regions around accreting high-mass stars
hints that these maser models may apply in both environments.

Finally, we note that no other masers detected in the region coincide with the
CS masers.  The closest OH and water masers \citep{Fish2007a,Sato2010a} are
separated by $>0.2\arcsec$ ($>1000$ AU).  There are some \methanol maser spots
with positions close to W51e2e's core ($\sim350$ AU, 0.07\arcsec) and with the
CS masers at a velocity $\sim60$ \kms, but they are not co-located
\citep{Etoka2012a,Surcis2012a}.  No other known masers are clearly emitted from
the same positions as the CS masers.  Appendix \ref{sec:extrafigs} shows maser
locations overlaid on Figure \ref{fig:overlay}.

\section{Conclusions}
We have detected two emission lines from the J=2-1 and J=1-0 transitions of CS
with high brightness temperature ($\geq$6800~K) indicating that these lines are masers.
This is the first reported detection of maser emission from CS.
While the HMYSO W51 e2e exhibits these masers, several neighboring HMYSOs that
are similar in luminosity, core mass, and apparent evolutionary stage do not.
The presence of two CS masers from different rotational states at different
velocities and locations in one source, W51 e2e, combined with the absence of
either of these masers in the other HMYSOs in the region, W51 IRS2/North and W51
e8, suggests that there is something unique about W51 e2e's radiation field,
geometry, or chemistry that promotes CS maser formation.

These CS masers join a growing list of rarely-detected maser transitions
that may trace a unique phase in the formation of high-mass protostars.
Like the NH$_3$, H$_2$CO, and SiO masers, there are only 1-10 known sources of
each of these masers.  Curiously, W51 IRS2/North, a known host of NH$_3$ and SiO masers,
does not exhibit any CS maser emission.  Other HMYSOs should be searched for
masers in these transitions to determine how common they are and precisely
what evolutionary stage they trace.

\textbf{Acknowledgements}
We thank Todd Hunter for providing literature references on CS masers and
Lorant Sjouwerman for discussion about a lack of CS masers in AGB stars.
We thank the anonymous referee for a helpful review.

This paper makes use of the following ALMA data sets:
ADS/JAO.ALMA\#2013.1.01596.S and 2017.1.00293.S. ALMA is a partnership of ESO
(representing its member states), NSF (USA) and NINS (Japan), together with NRC
(Canada), MOST and ASIAA (Taiwan), and KASI (Republic of Korea), in cooperation
with the Republic of Chile. The Joint ALMA Observatory is operated by ESO,
AUI/NRAO and NAOJ.  The National Radio Astronomy Observatory is a facility of
the National Science Foundation operated under cooperative agreement by
Associated Universities, Inc.

\textbf{Software}
This paper used \texttt{astropy}
\citep{Astropy-Collaboration2013a,Astropy-Collaboration2018a},
\texttt{pyspeckit} \citep{Ginsburg2011c}, and CASA \citep{McMullin2007a}.

\appendix
\section{Velocity check}
We carefully checked the velocity measurements in the VLA data, since an offset
of a few channels is comparable to the Earth's motion and we are using a
continuum band to infer velocity information.  The observations we analyze were
taken on December 26 and 30, 2016 and January 7, 2017.  On these dates, the
topocentric-to-barycentric velocity corrections are -9.0, -7.4, and -4.1 \kms,
respectively.  The barycentric-to-LSR correction toward W51 is 16.5 \kms.  The
observed CS J=1-0 peak is in channel 24 (zero-indexed) of spectral window 26,
which has channel 0 at  48957.667, 48957.932, and 48958.478 MHz, respectively,
for each of the three observations.  The channel width is 1 MHz.  The LSR
velocity of the CS J=1-0 peak intensity channel is at 64.3 \kms on all three
dates.  This value is consistent with our reported velocity of
$v_{LSR}(\mathrm{CS~J=}1-0) = 65.5 \pm 0.7 \kms$.

No such sanity check is needed for the ALMA data, since the channel maps
clearly show the outflow at appropriate velocities with morphology that matches
the SiO outflow seen at both 1 mm and 3 mm.

\section{Variability check}
We also checked for variability in the CS 1-0 maser, and found no evidence for
variability across four observing epochs (2016-10-02, 2016-12-26, 2016-12-30,
2017-01-07; the former was not included in our merged data set because we were unable
to perform bandpass calibration).  The position remained constant and the flux
remained consistent to within single-epoch observing errors.

\section{Additional figures showing other maser locations}
\label{sec:extrafigs}
\begin{figure*}[htp]
    \includegraphics[width=0.33\textwidth]{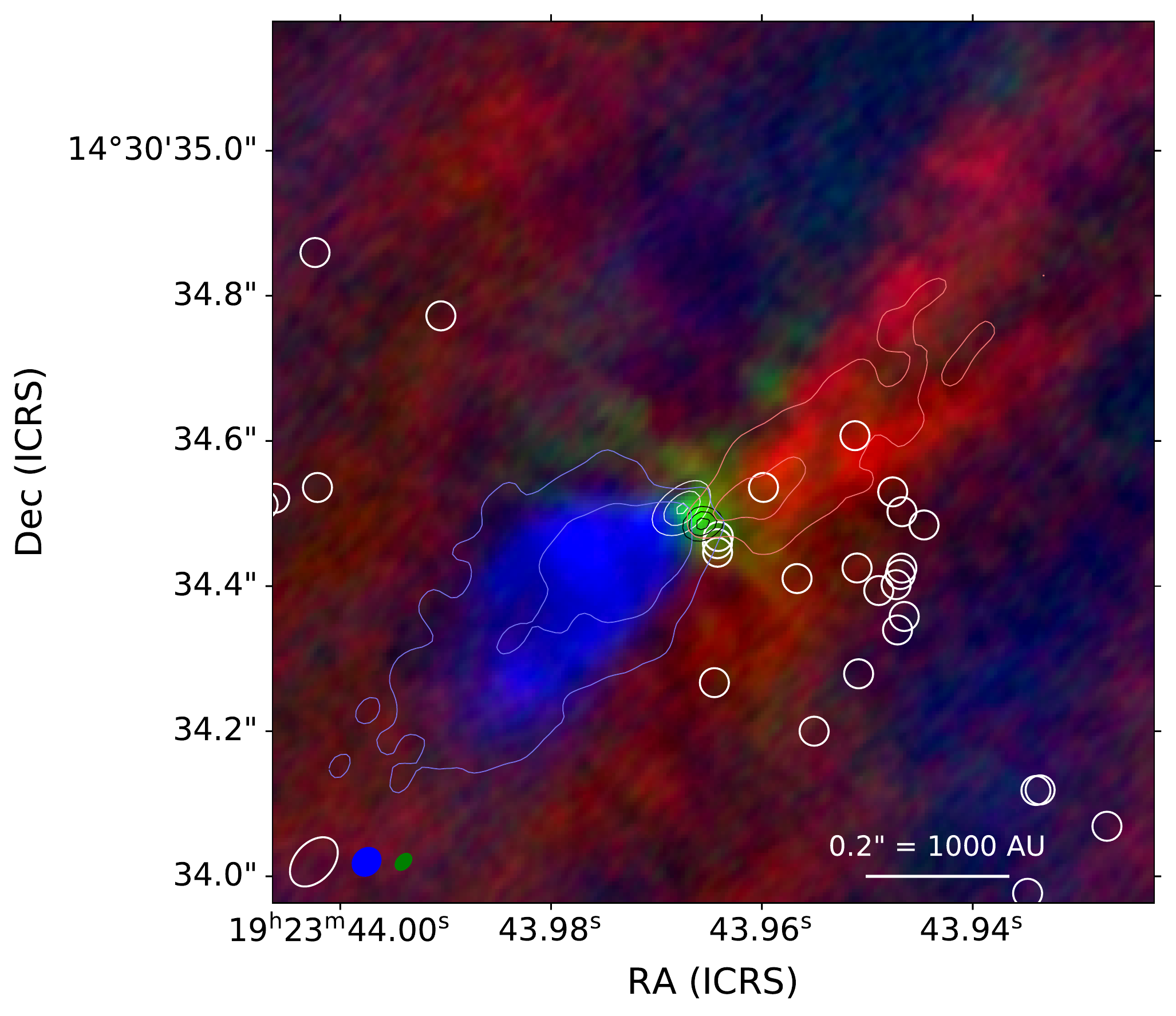}
    \includegraphics[width=0.33\textwidth]{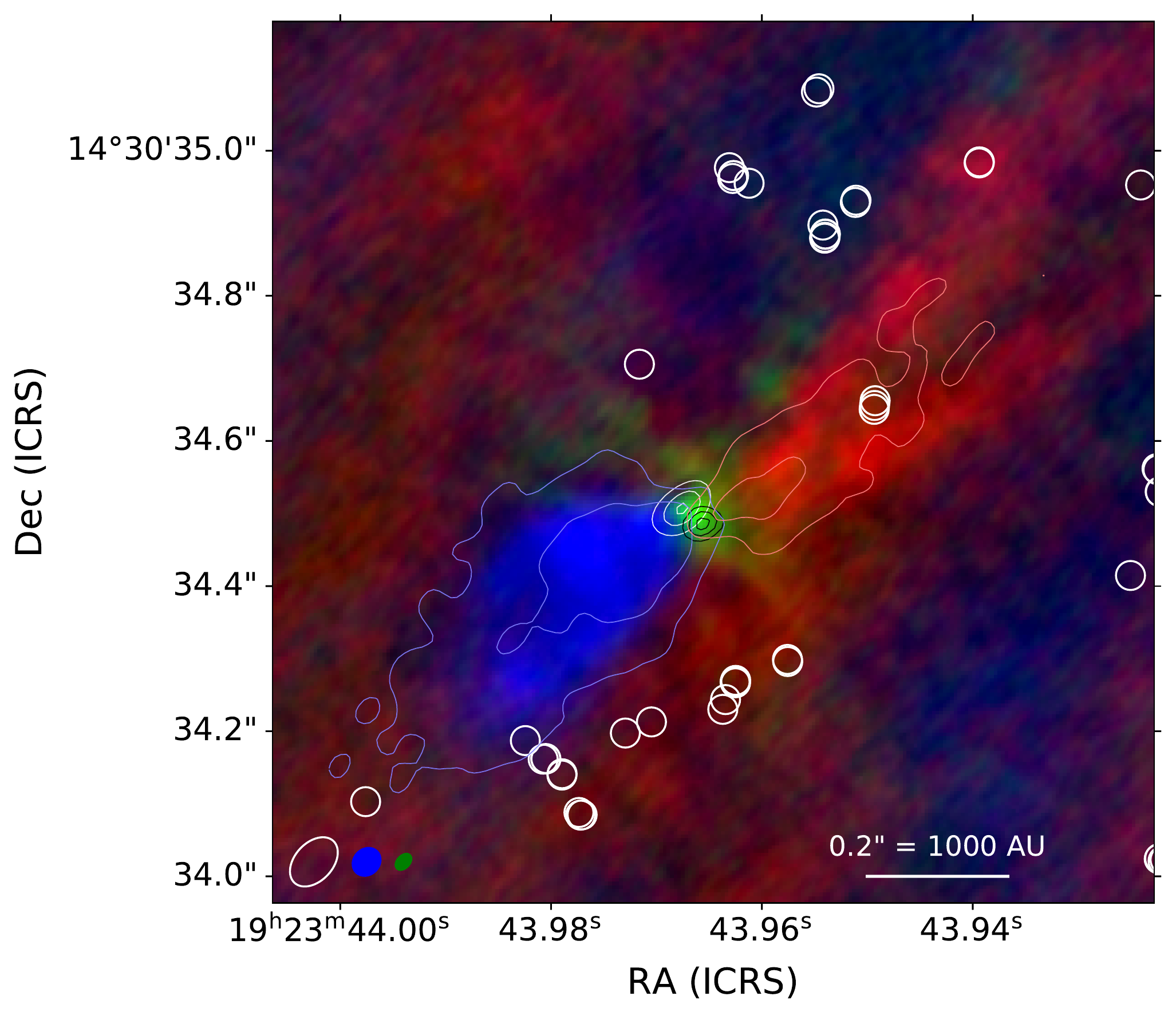}
    \includegraphics[width=0.33\textwidth]{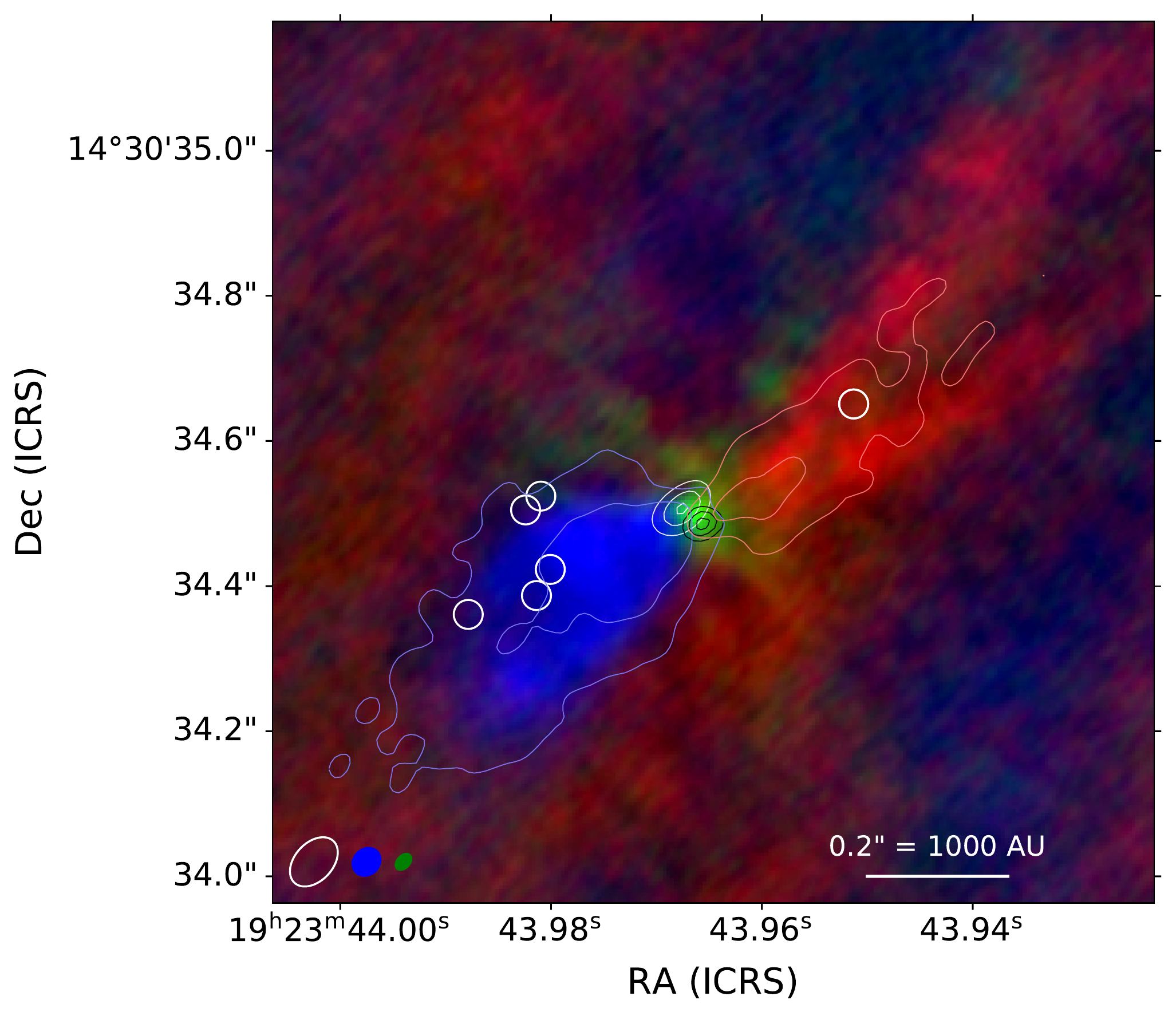}
    \caption{Figure \ref{fig:overlay} reproduced with (a) \methanol maser spots
    from \citet{Etoka2012a}, (b) OH maser spots from \citet{Fish2007a}, and (c)
    \water maser spots from \citet{Sato2010a} overlaid.
    }
    \label{fig:maseroverlay}
\end{figure*}

We show Figure \ref{fig:overlay} with maser positions from \methanol, OH, and
\water (Fig.  \ref{fig:maseroverlay}) in this Appendix.  These figures
highlight that the conditions to excite each of these masers is unique.

\end{document}